\documentclass[twocolumn,superscriptaddress,showpacs,aps,amsmath,amssymb,prb]{revtex4}

\usepackage{soul}

\usepackage[english]{babel}

\usepackage{braket}
\usepackage{tikz}
\usepackage{relsize}
\usepackage{framed}
\DeclareMathOperator{\Tr}{\text{Tr}}

\usepackage{bm}
\usepackage{graphicx,color}
\usepackage{hyperref}

\usepackage{amsfonts}
\usepackage{mathrsfs}

\newcommand{\ti}{\Tilde}
\newcommand{\nl}{\nonumber \\}
\newcommand{\nla}{\nl&\quad}

\newcommand{\up}{\uparrow}
\newcommand{\down}{\downarrow}

\newcommand{\Sec}[1]{Sec.\;\ref{#1}}
\newcommand{\App}[1]{Appendix\;\ref{#1}}
\newcommand{\be}{\begin{equation}}
\newcommand{\ee}{\end{equation}}
\newcommand{\bea}{\begin{eqnarray}}
\newcommand{\eea}{\end{eqnarray}}
\newcommand{\G}{\mbox{\tiny $\Sigma$}}
\newcommand{\B}{\mbox{\tiny B}}
\newcommand{\s}{\mbox{\tiny S}}

\newcommand{\bsube}{\begin{subequations}}
\newcommand{\esube}{\end{subequations}}
\newcommand{\Eq}[1]{Eq.\,(\ref{#1})}
\newcommand{\Eqs}[1]{Eqs.\,(\ref{#1})}
\newcommand{\Fig}[1]{Fig.\,\ref{#1}}

\newcommand{\dg}{\dagger}
\newcommand{\la}{\langle}
\newcommand{\ra}{\rangle}

\begin{document}

\title{Non-Markovian correlation functions for open quantum systems}

\author{Jinshuang Jin} \email{jsjin@hznu.edu.cn}
\affiliation {Institute of Nanotechnology, Karlsruhe Institute of Technology (KIT), 76344 Karlsruhe, Germany}
\affiliation{ Department of Physics, Hangzhou Normal University,
  Hangzhou 310036, China}
\affiliation {Institut f\"ur Theoretische Festk\"orperphysik,
      Karlsruhe Institute of Technology, 76131 Karlsruhe, Germany}

 \author{Christian Karlewski}
 \affiliation {Institute of Nanotechnology, Karlsruhe Institute of Technology (KIT), 76344 Karlsruhe, Germany}
\affiliation {Institut f\"ur Theoretische Festk\"orperphysik,
      Karlsruhe Institute of Technology, 76131 Karlsruhe, Germany}
      
\author{Michael Marthaler}
\affiliation {Institut f\"ur Theoretische Festk\"orperphysik,
      Karlsruhe Institute of Technology, 76131 Karlsruhe, Germany}

\date{\today}

\begin{abstract}
Beyond the conventional quantum regression theorem,
a general formula for non-Markovian correlation functions 
of arbitrary system operators both in the time- and frequency-domain is given. 
We approach the problem by transforming the conventional time-nonlocal master equation
 into dispersed time-local equations-of-motion.
The validity of our approximations is discussed and we find that the 
non-Markovian terms  have to be included 
for short times. While calculations of the 
density matrix at short times suffer from the initial value problem, 
a correlation function has a well defined initial state.  
 The resulting formula for the non-Markovian correlation function
 has a simple structure and is as  convenient in its application as the conventional quantum
  regression theorem for the Markovian case.
 For illustrations, we apply our method to
 investigate the spectrum of the current fluctuations of
  interacting quantum dots contacted with two electrodes. 
The corresponding  non-Markovian characteristics are demonstrated.

\end{abstract}

\maketitle

\section{Introduction}
\label{intro}

Open quantum systems, which are of great importance in many fields of physics,
refer to a quantum system of primary interest coupled to an
environment often called reservoir or bath. \cite{Car93,Scu97,Gar00,Bre02,Wei08,Saptsov_long_paper}
The composite Hamiltonian ($H_{\rm tot}$), in general, 
includes the system ($H_{\s}$), the bath ($H_{\B}$), and the coupling between the system
and the bath ($H'$), i.e., $H_{\rm tot}=H_{\s}+H_{\B}+H'$.
It is well-known that the system
 is described by the reduced density operator, $\rho(t)\equiv{\rm tr}_B[\rho_{\rm tot}(t)]$,
i.e., the partial trace of the total density operator $\rho_{\rm tot}$ over the bath
space. The corresponding
dynamics is determined by the master equation,  
\be\label{ME0}
 \dot\rho(t) = -i[H_{\s},\rho(t)]
  - \int_{t_0}^t\!\!{\mathrm d}\tau
   \Sigma(t-\tau)\rho(\tau),
\ee
  where the effect of the bath is described by the second term with
  the self-energy $\Sigma(t-\tau)$. It contains the memory effect in principle even for weak 
  system-reservoir coupling.
  \Eq{ME0} is thus the so-called time-nonlocal (TNL) master equation describing
   non-Markovian dynamics.

  As long as one knows the reduced density operator $\rho(t)$, a single-time 
  expectation value of an arbitrary physical observable of the system, e.g., $\hat O$, is simply obtained via
  $\la O(t) \ra={\rm Tr}[\hat O\rho(t)]$. However, it is not as easy to
  calculate two- or multiple-time correlation functions. 
Except for a small number of exactly solvable systems, 
convenient calculation of the correlation function is possible 
using the well-known quantum regression theorem 
which is valid in the  Born-Markovian approximation.
\cite{Lax632342,Lax67213,Car93,Scu97,Gar00,Bre02,Wei08} 
It is not valid any more for non-Markovian cases,
\cite{For96798,For99144,For00451,Lax00463,Wei08}
due to memory effects which
break the time translation invariance of the 
non-Markovian propagator.

The calculation of non-Markovian correlation functions for arbitrary system operators
is a challenge and long-standing problem. 
Stimulated not only by fundamental interest but also
by great demand because of the rapid
progress in experiments which are able to access non-Markovian 
effects, \cite{Ber06759,Win09207403,Ulr11247402} there are many efforts to address this issue.
For example, 
by using the stochastic Schr\"odinger equation approach and the Heisenberg equation, 
\cite{Alo05200403,Alo07052108,Veg06022102}
rather than master equation,
the method derived by Alonso and de Vega \cite{Alo05200403,Alo07052108,Veg06022102}
is valid for zero-temperature environments and/or Hermitian system-environment coupling operators.
Based on a generalized Born-Markov approximation, Budini \cite{Bud0851} derived a quantum regression theorem 
which is applicable for non-Markovian Lindblad equations.
Recently, Goan {\it et al}., \cite{Goa11115439,Goa11124112} developed a 
scheme for the calculation of two-time correlation functions 
of the system operators with memory effects in terms of the ``time-convolutionless" master equation.
Additionally some specific systems  
have been analyzed \cite{Shn0211618,Kac03035301,Joh02046802}.

In this work, we aim to derive a general formula for 
 non-Markovian correlation 
functions of arbitrary system operators 
in terms of the TNL-ME \Eq{ME0}. We will consider weak system-reservoir coupling but
short time-scales where non-Markovian effects should dominate. Later we will also analyze the relevant time scales in more detail. 
By using the fluctuation dissipation theorem of the correlator
and introducing the auxiliary density operator in the frequency domain denoted by $\phi^{\pm}(\omega,t)$,
it is easy to transform the TNL-ME \Eq{ME0} into an equivalent set of coupled time-local
 equations-of-motion (for short TL-EOMs, expressed in \Eq{TL-EOM}),\cite{Jin08234703}
 i.e., $\dot {\vec\rho}(t)={\bf{\Lambda}}\vec{\rho}(t)$,
   in the enlarged vector space with $\vec{\rho}(t)
\equiv\left[\rho(t),\phi^+(\omega,t),\phi^-(\omega,t) \right]^T$. 
 The corresponding propagator in this enlarged vector space satisfies time-translation invariance
 and accordingly the correlator can be treated  similar to the Markovian case.
 The resulting equation takes a form very similar to the quantum regression theorem
 in a larger space,
\be\label{ABt0_first}
\la \hat A(t)\hat B(0)\ra\! 
=\!{\rm Tr}\left\{\hat A \,
\big[\vec\Pi(t,0)\vec{\rho}_B(0)\big]\right\},
\ee 
The most important feature of this approach is the initial condition $\vec{\rho}_B(0)=B\vec{\rho}(0)$,
where $\vec{\rho}(0)$ is the density matrix which has been time evolved from a initial time $t_0\rightarrow -\infty$. We 
assumed that system and reservoir decouple at the initial  
 time $t_0\!\rightarrow\!-\infty$ which is the standard assumption for the  TNL-ME. In the following
 we will derive equation (\ref{ABt0_first}) and we will discuss in detail 
 the range of validity using the diagrammatic expansion on Keldysh contour \cite{Sch9418436}.
This full non-Markovian description is  applicable for both,
fermionic and bosonic systems.
As an example, we will discuss non-Markovian effects of the electronic reservoir on the current-fluctuation 
 spectrum in quantum dots.

 The paper is organized as follows.
In \Sec{thnmkME}, we first introduce the conventional TNL--ME
for weak system-reservoir coupling and then outline the equivalent TL-EOMs.
Based on TL-EOMs, we get the formulas for the two-time non-Markovian correlation functions both in the time-domain 
and in the frequency-domain in \Sec{thnmkcf}.
We then implement the proposed scheme to 
study the current-fluctuation spectra of the electron tunneling through
quantum dots in \Sec{thexa}.
Finally, we conclude in \Sec{sum}.

\section{Non-Markovian master equation: TNL-ME versus TL-EOMs}
\label{thnmkME}

\subsection{Time non-local master equation}

The reservoir with infinite degrees of freedom 
 is described by the non-interacting Hamiltonian $H_{\B}=\sum_{ k} \epsilon_{ k}
  c^{\dg}_{ k}c_{ k}$ with the creation (annihilation) operator $c^\dg_k$ ($c_k$).
The coupling Hamiltonian between the system and the bath, in general, is given by
\be\label{Hprime0}
  H'=  Q^+ F^-+ F^+Q^-, 
 \ee
 where $(Q^+)^\dg= Q^-$ and $(F^-)^\dg= F^+$, with 
the operator of the central system  $ Q^{\pm}$ and
      the operator of the bath $ F^{\pm}$. The coupling operator of the bath is defined as 
    $F^-=\sum_k t_{ k} c_k$ and contains the coupling coefficients $t_{ k}$.
The result will be generalized for the case of coupling to multiple reservoirs 
in the \App{app_Hmul}.
The Hamiltonian of the small system is composed of the corresponding creation ($a^\dg_\mu$) and annihilation ($a_\mu$) operators,
i.e., $H_{\s}\equiv H_{\s}(a_\mu,a^\dg_\mu)$ which could include many-body interaction terms.

\begin{figure}
\centerline{\includegraphics*[width=1.0\columnwidth,angle=0]{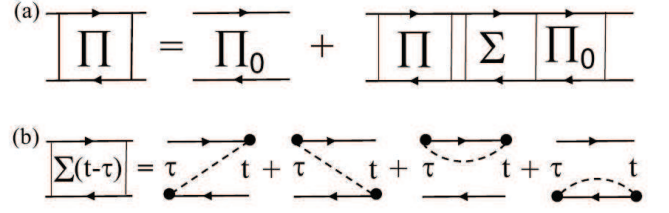}}  
\caption{ 
The diagrams of (a) Dyson equation and (b) the self-energy with the lowest-order contributions. \cite{Sch9418436} 
In (a), $\Pi_0$ is the free propagator with $\Pi_0(t,t_0)=e^{-i{\cal L_{\s}}(t-t_0)}$,
and the dashed line in (b) is the correlation 
function $C^{(\pm)}(t-\tau)$ of the bath expressed in \Eq{ct}.}
  \label{figdyson}
\end{figure}

Assuming weak system-bath coupling and performing Born but without
Markovian approximation,
the self-energy for the expansion up to second-order 
of the coupling Hamiltonian is expressed as
$\Sigma(t-\tau)\rho(\tau)=\big\la[H'(t),e^{-iH_{\s}(t-\tau)}[H'(\tau),\rho(\tau)]e^{iH_{\s}(t-\tau)}]\big\ra_{\B}$
in the $H_{\B}$-interaction picture.
The corresponding diagram is schematically shown in \Fig{figdyson}(b).
Here, $\la \cdots\ra_{\B}$ stands for the statistical average
over the bath in thermal equilibrium.
The explicit formalism for the self-energy in \Eq{ME0} thus reads, 
 \be\label{selft}
 \Sigma(t-\tau)\bullet=\big[Q^{\mp},
\Pi_0(t,\tau){\cal C}^{(\pm)}_Q(t-\tau) \bullet \big],
  \ee
where we introduced the free propagator 
$\Pi_0(t,\tau)\equiv e^{-i{\cal L}_{\s}(t-\tau)}$ with
  ${\cal L}_{\s} \bullet=[H_{\s},\bullet]$, 
 and
 \be\label{calCt}
 {\cal C}^{(\pm)}_Q(t) \bullet\equiv C^{(\pm)}(t)Q^\pm\bullet-\bullet C^{(\mp)\ast}(t)Q^\pm,
 \ee
 with the correlation function of the bath,
\begin{align}\label{ct}
    C^{(\pm)} (t-\tau)
  & = \la F^\pm (t) F^{\mp} (\tau)  \ra_{\B}.
  \end{align}
Consequently, the TNL-ME \Eq{ME0} is explicitly given by ($\hbar=1$), \cite{Jin11053704,Yan05187,Mak01357}
\be\label{ME1}
\dot\rho(t)\! =\! -i{\cal L}_{\s}\rho(t)-\sum_{+,-}
  \int_{t_0}^t\!\!{\mathrm d}\tau \big[Q^{\mp},
  \Pi_0(t,\tau){\cal C}^{(\pm)}_Q(t-\tau) \rho(\tau) \big].
 \ee
 Note that \Eq{ME1} is derived 
 assuming initial decoupling at  $t_0\rightarrow-\infty$, 
$\rho_{\rm tot}(t_0)=\rho(t_0)\rho_{\B}$ and $\rho_{\B}$ is the equilibrium state of the bath.
The corresponding propagator $\Pi(t,t_0)$ for \Eq{ME1}( \Eq{ME0}) defined by 
 \be\label{rho_prop}
 \rho(t)=\Pi(t,t_0)\rho(t_0),
 \ee
satisfies the Dyson equation \cite{Sch9418436} 
as shown in Fig.\,\ref{figdyson} (a)
and does not satisfies time-translation invariance, i.e.
$\Pi(t,t_0)\neq \Pi(t,t_1)\Pi(t_1,t_0)$.
The conventional quantum regression theorem is thus  broken.

\subsection{Time-local equations-of-motion}

The key to the calculation of the non-Markovian correlation function, e.g.,
$\la \hat A(t)\hat B(\tau) \ra$, is how to expand the master equation to an extend space $\rho\rightarrow \vec{\rho}$
which preserves again time translation symmetry. This will allow us to case
the correlator into the form of the regression theorem \Eq{ABt0_first}.  

We adopt the multi-frequency-dispersed scheme \cite{Jin07134113,Jin08234703}  
and define the bath correlation function \Eq{ct} in the frequency-domain ($C^{(\pm)}(\omega)$) as
 \be \label{FDT}
    C^{(\pm)}(t)=
   \int^{\infty}_{-\infty}\!\!\frac{d\omega}{2\pi}\,
   e^{ \pm i\omega t}C^{(\pm)}(\omega),
  \ee
where $C^{(\pm)}(\omega)$ is directly related to the spectral density of the bath depending on the 
specific operator ($F^\pm$). Correspondingly, the Liouville operator of \Eq{calCt} in the
frequency-domain is [c.f.\Eq{FDT2}]
 \be\label{calCw}
 {\cal C}^{(\pm)}_Q(\omega) \bullet\equiv C^{(\pm)}(\omega)Q^\pm\bullet-\bullet C^{(\mp)\ast}(\omega)Q^\pm.
 \ee

Furthermore, we introduce the auxiliary density operators
in the frequency-domain defined by
\begin{align}\label{phi0}
 \phi^{\pm}(\omega,t) &= 
  -i \int_{t_0}^t\!\!{\mathrm d}\tau
  e^{-i({\cal L}_{\s}\mp\omega)(t-\tau)}{\cal C}^{(\pm)}_Q(\omega) \rho(\tau),
 \end{align}
 which means $\phi^{\pm}(\omega,t_0)=0$ is applicable for the initially
 decoupled system-reservoir with $t_0\!\!\rightarrow\!\!-\infty$ as we mentioned above.
Taking the time derivatives of the auxiliary density operators, it is easy to 
recast TNL-ME of \Eq{ME1}( \Eq{ME0}) in the form
\bsube\label{TL-EOM}
\begin{align}
 &\dot\rho(t)
 =-i{\cal L}_{\s}\rho(t)-i\sum_{+,-}\int \frac{d\omega}{2\pi}
 \big[ Q^{\mp},\phi^{\pm}(\omega,t)\big],
\label{rho0t}
\\
&\dot\phi^{\pm}(\omega,t)
=-i({\cal L}_{\s}\mp\omega)\phi^{\pm}(\omega,t)
  -i{\cal C}^{(\pm)}_Q(\omega) \rho(t),
 \label{rho1t}
\end{align}
\esube
which are the so-called
time-local equations-of-motion (TL-EOMs) 
 due to the involved time-independent dissipative coefficients.
TL-EOMs \Eq{TL-EOM} is the lowest-tier truncation of hierarchical equation of motion \cite{Jin08234703}
which has the linearity of the hierarchical Liouville space as demonstrated in Ref.\,\onlinecite{Wan13035129}.

Introducing a vector composed of
the reduced density operator and auxiliary density operators, i.e.,
 \be\label{vec0}
\vec{\rho}(t)
\!\equiv\!\left[\rho(t),\phi^{+}(\omega,t) ,\phi^{-}(\omega,t)\right]^T,
\ee
the TL-EOMs \Eq{TL-EOM} can then be further
 compacted with
\be\label{vecrho}
\dot {\vec\rho}(t)={\bf{\Lambda}}\vec{\rho}(t).
\ee
Here according to \Eq{TL-EOM}, ${\bf{\Lambda}}$ can be formally written as
\begin{align}\label{Lambda}
{\bf{\Lambda}}&=
\left(\begin{array}{ccc}
-i{\cal L}_{\s}& -i\int \frac{d\omega}{2\pi} {\cal Q}^+ &-i\int \frac{d\omega}{2\pi} {\cal Q}^- 
\\
-i{\cal C}^{(+)}_Q(\omega)&-i({\cal L}_{\s}-\omega) & 0 
\\
-i{\cal C}^{(-)}_Q(\omega)& 0 & -i({\cal L}_{\s}+\omega) 
\end{array}
\right),
\end{align}
where we introduced ${\cal Q}^{\pm}\bullet=[Q^{\pm},\bullet]$.

Apparently, \Eq{vecrho}
leads to
$ {\vec\rho}(t)=\vec\Pi(t,t_0)\vec{\rho}(t_0)$ with 
$\vec\Pi(t,t_0)=e^{\bf{\Lambda}(t-t_0)}$. 
In this vector space, the propagator satisfies
the time-translation invariance, i.e., $\vec\Pi(t,t_0)=
\vec\Pi(t,\tau)\vec\Pi(\tau,t_0)$
and the correlation function \ref{ABt0_first} can be calculated straightforwardly 
in a form similar to the  Markovian case based on the quantum regression theorem.


\section{Non-Markovian correlation function}
\label{thnmkcf}

\subsection{Two-time correlation function}
Using the vector of $\vec\rho(t)$ defined in \Eq{vec0}, a single-time expectation value of the system operator $\hat A$ can be
expressed as
 \begin{align}\label{At}
\la \hat A(t)\ra
={\rm Tr}[\hat A \rho(t)]
={\rm Tr}\big\{\hat A\, [\vec\Pi(t,t_0)\vec\rho(t_0)]\big\},
\end{align}
with the initial condition being that system and reservoir decouple ($\phi^{\pm}(\omega,t_0\!\!\rightarrow\!\!-\infty)=0$),
  i.e., $\vec\rho(t_0)=\{ \rho(t_0),0, 0 \}$.
Since time-translation invariance of the propagator has been restored, 
we can now follow the derivation of the Markovian correlation function based on the quantum regression theorem. \cite{Bre02}
Therefore the non-Markovian two-time correlation function in the vector space is given by, 
\be\label{ABt0}
\la \hat A(t)\hat B(0)\ra\!
\!=\!{\rm Tr}\left\{\hat A \,
\big[\vec\Pi(t,\tau)\hat B \vec\rho(0)\big]\right\},
\ee 
where the components of the density operators in the vector are now
 \be
\hat{\vec{\rho}}(t)
\!\equiv\!\left[\hat\rho(t),\hat\phi^{+}(\omega,t) ,\hat\phi^{-}(\omega,t)\right]^T.
\ee
The initial state at $t=0$ is given by 
\be\label{rhotau}
\hat B\vec{\rho}(0)=\{ \hat B\rho(0),\hat B \phi^+(\omega,0),\hat B \phi^-(\omega,0)\},
\ee
where $\vec{\rho}(0)$ is the density matrix 
which has been time evolved from a initial time $t_0\rightarrow -\infty$.

A similar equation  
has been derived using  
linear response theory in Ref.\,\onlinecite{Wan13035129}.
In this case all high-order contributions in the self-energy have been considered and
 one should keep all hierarchical EOMs for the numerical calculation of 
 the correlation function. \cite{Jin08234703,Wan13035129}

In this paper we consider the lowest-order contribution for weak system-reservoir coupling.
In terms of eqs. \Eqs{ABt0}-(\ref{rhotau})
together with the TL-EOMs \Eq{TL-EOM},
 (see \App{app_cf} for the detail),
we finally get
\begin{widetext}
\begin{align}\label{CABt}
\la \hat A(t)\hat B(0)\ra \!=\!
{\rm Tr}\left[\hat A\,\Pi(t,0)\hat B\,\rho(0)\right]
\!-\!\sum_{+,-}\int^t_{0}\!\!dt_2\int^{0}_{t_0}\!\!dt_1 {\rm Tr}\Big\{\hat A\,\Pi(t,t_2)
\big[Q^{\mp},\Pi_0(t_2,0)\hat B \,\Pi_0(0,t_1)
{\cal C}^{(\pm)}_Q(t_2-t_1)\rho(t_1)\big]\Big\},
\end{align}
 \end{widetext}
with the steady-state $\rho(0)=\Pi(0,t_0)\rho(t_0)=\bar\rho$ and $\rho(t_1)=\Pi(t_1,t_0)\rho(t_0)=\bar\rho$
for $t_0\!\rightarrow\!-\infty$. 
Compared to the Markovian correlation function, the second modification in \Eq{CABt} arises from the
memory effect and it is the 
vertex corrections which will be further illustrated in the coming subsection based on a diagrammatic representation.

Similarly, it is easy to get the NMK-CF of $\la \hat A(0)\hat B(t)\ra$
as expressed in \Eq{CBAt}. 
In this vector space, 
Non-Markovian multiple-time correlation functions 
 can be calculated by $\la \hat A(t)\hat C(\varsigma)\cdots \hat B(0)\ra 
={\rm Tr}\left\{\hat A \,
\big[\vec\Pi(t,\varsigma)\hat C\vec\Pi(\varsigma,\tau)\cdots\hat B \vec\rho(0)\big]_1\right\}$.

\subsection{Spectrum of the two-time correlation function}

\label{theo-spec}

Since the widely measured quantity in experiments is 
the spectrum of the correlation function (in Fourier space, i.e.,
${\cal F} [f(t)]\equiv\int^\infty_{-\infty}dt\,e^{i\omega  t}f(t)$),
  we will now calculate
the spectrum for the stationary two-time correlation function,
\begin{align}
S_{AB}(\omega)
\!\equiv\! {\cal F} [\la A(t)B(0)\ra]
\! =\!2{\rm Re}\big\{L [\la A(t)B(0)\ra]\},
\end{align}
where the last identity 
is assumed to be $\hat A^\dg=\hat B$, and
$L [\la A(t)B(0)\ra]$ is the Laplace transformation defined by 
$L [f(t)]\equiv\int^\infty_{0}dt\,e^{i\omega  t}f(t)$.
Based on either \Eq{CABt} directly or \Eq{ABt0},
 we finally obtain (for the detail derivation see the \App{app_cf}),
\begin{align}\label{CABw}
&L [\la A(t)B(0)\ra] 
={\rm Tr}[\hat A\, \widetilde\Pi(\omega)\hat B\bar\rho]\!-\!
\frac{i}{\omega}{\rm Tr}\Big\{\hat A\,\widetilde\Pi(\omega)
\nl&\quad\quad\quad\times
\!\sum_{+,-}\!\big[Q^{\mp}\!\!,\hat B (\widetilde{\cal C}^{(\pm)}_Q({\cal L}_{\s},0)
\!-\!\widetilde{\cal C}^{(\pm)}_Q({\cal L}_{\s},\omega))\bar\rho\big] \Big\},
\end{align}
where $\widetilde\Pi(\omega)$ and $\widetilde{\cal C}^{(\pm)}_Q({\cal L}_{\s},\omega)$ are the
counterparts in the frequency domain by Laplace transformation of $ \Pi(t,t_0)$ [c.f.\Eq{rho_prop}]
and $ e^{-i{\cal L}_{\s}t}{\cal C}^{(\pm)}_Q(t)$ [c.f.\Eq{calCt}], respectively.
Explicitly, they are given by
\bsube
\begin{align}
\widetilde\Pi(\omega)&=\big[i({\cal L}_{\s}-\omega)-\widetilde\Sigma(\omega)\big]^{-1},
\end{align}
and
 \begin{align}
\widetilde {\cal C}^{(\pm)}_Q({\cal L}_{\s},\omega)\bullet\!=\!
\Big[\widetilde{C}^{(\pm)}(\omega-{\cal L_{\s}})Q^\pm\!\bullet
-\bullet\!\widetilde C^{(\mp)\ast}({\cal L}_{\s}-\omega)Q^\pm\Big],
\end{align}
\esube
with the frequency-domain of the self-energy [\Eq{selft}]
\bsube
\be
\widetilde\Sigma(\omega)\bullet=\sum_{+,-}
[{ Q}^{\mp}, \widetilde {\cal C}^{(\pm)}_{Q }({\cal L}_{\s},\omega)\bullet],
\ee
and the frequency-domain of the bath correlation [c.f.\Eq{ct} and \Eq{FDT}] 
\begin{align}\label{tcw0}
   \widetilde C^{(\pm)} (\omega)
&=\int^\infty_{-\infty}\frac{d\omega'}{2\pi}
\frac{i}{\omega\pm\omega'+i0^+}C^{(\pm)}(\omega').
\end{align}
\esube
It can be further give by
\be
\widetilde C^{(\pm)} (\omega)=\frac{1}{2}\left[C^{(\pm)}(\mp \omega)+i\Lambda^{(\pm)} (\omega)\right],
\ee
where the real part $C^{(\pm)}( \omega)$ is described by \Eq{FDT} and the imaginary part
is
$\Lambda^{(\pm)} (\omega)
\equiv{\cal P}\int^\infty_{-\infty}\frac{d\omega'}{2\pi}
\frac{1}{\omega\pm\omega'}C^{(\pm)} (\omega)$,
with ${\cal P}$ denoting the principle value of the integral.

Similarly, the spectrum of the correlation function of $\la \hat A(0)\hat B(t)\ra$, 
 is given by \Eq{CBAw}.
The formulas  \Eq{CABt} and \Eq{CABw} are the main contributions of the present work for the calculation of
the non-Markovian two-time correlation function of arbitrary system operators  
(denoted by $\hat A$ and $\hat B$) in time-domain and the frequency-domain, respectively.

Note that for often relevant Hermitian coupling operator, i.e., $H'=QX$ [c.f.\Eq{Hprime0}] with $Q^\dg=Q$ and $X=F^++F^-$,
\Eq{CABt} and \Eq{CABw} is simplified to, 
\bsube\label{CABtwh}
\begin{align}\label{CABth}
&\la \hat A(t)\hat B(0)\ra \!=\!
{\rm Tr}\left[\hat A\,\Pi(t,0)\hat B\,\rho(0)\right]
\!-\!\sum_{+,-}\int^t_{0}\!\!dt_2\int^{0}_{t_0}\!\!dt_1
\nl&
\times\!
 {\rm Tr}\Big\{\hat A\,\Pi(t,t_2)
\big[Q,\Pi_0(t_2,0)\hat B \,\Pi_0(0,t_1)
{\cal C}_Q(t_2-t_1)\rho(t_1)\big]\Big\},
\end{align}
 and
\begin{align}\label{CABwh}
&L [\la A(t)B(0)\ra]
={\rm Tr}[\hat A\, \widetilde\Pi(\omega)\hat B\bar\rho]\!-\!
\frac{i}{\omega}{\rm Tr}\Big\{\hat A\,\widetilde\Pi(\omega)
\nl&\quad\quad\quad\times
\!\big[Q,\hat B (\widetilde{\cal C}_Q({\cal L}_{\s},0)
\!-\!\widetilde{\cal C}_Q({\cal L}_{\s},\omega))\bar\rho\big] \Big\},
\end{align}
\esube
respectively,
where we used $Q^-=Q^+=Q$ and $\widetilde{\cal C}_Q=\widetilde{\cal C}^{+}_Q+\widetilde{\cal C}^{-}_Q$
(${\cal C}_Q={\cal C}^{+}_Q+{\cal C}^{-}_Q$).

\subsection{Discussion and comments}

Now we are in the position to discuss the applicability and the range of validity of the
present NMK-CF formula \Eq{CABt} (or \Eq{CABw}). 
For convenience, we consider the coupling Hamiltonian $H'=QX$, with the operator of the bath $X$ containing
the coupling coefficient $g$, as an example. The resulting formula for non-Markovian correlation function
is given by \Eq{CABtwh}. The bath correlator in \Eq{ct} is recast to
\be\label{cth}
 C (t-\tau)
 = \la X^{\dg} (t) X (\tau)  \ra_{\B}
 =\sum^\infty_{m=0} \eta_m e^{-\gamma_m t},
 \ee
where the second identity is written as a parametric decomposition, \cite{Mei993365, Xu05041103}
with $\eta_m\propto g^2$ and $1/\gamma_m$ representing the correlation time of the bath or the so-called the memory time. 
For the present considered weak system-reservoir coupling, 
the self-energy expressed in \Eq{selft} contains the lowest-order contribution (or the first-order contraction 
by Wick theorem), e.g.,
$\Sigma(t)\propto C(t)$. Let the index ``$l$'' represent the number of contractions.
The lowest-order contribution considered in the self-energy means $l=1$.

\begin{figure}
\centerline{\includegraphics*[width=1.0\columnwidth,angle=0]{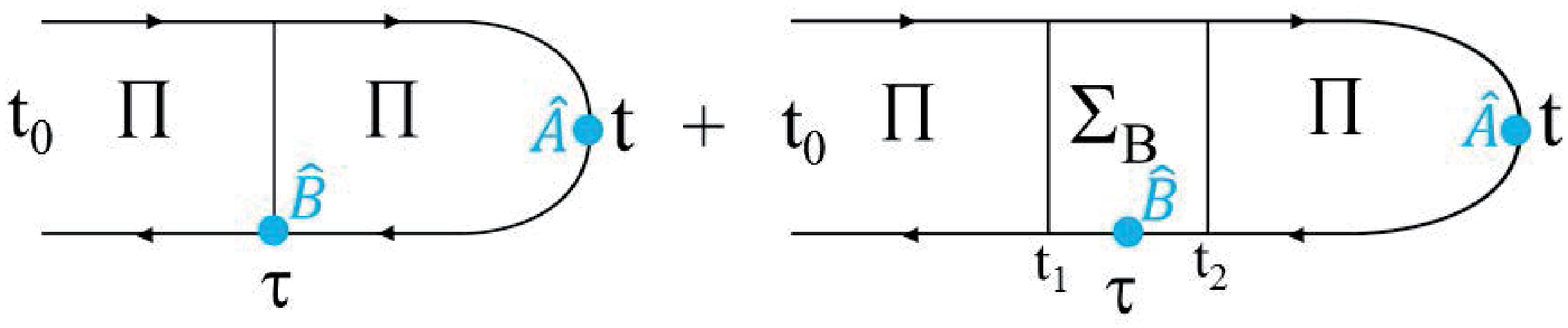}}  
\caption{ (Color online)
The diagram of the two-time correlation function
of the system operators $\la \hat A(t)\hat B(\tau)\ra$.
$\Sigma_B$ is the self-energy for vertex corrections 
}
  \label{figvertex}
\end{figure}

In the following discussion we will not limit us to $l=1$, but consider arbitrary number of the contractions such that the
self-energy is exact. Based on the diagrammatic technique,\cite{Sch9418436} we can obtain (see \App{app_dia_CAB} for details),
\begin{align}\label{CABt-dia}
&\la \hat A(t)\hat B(0)\ra =
{\rm Tr}\left[\hat A\,\Pi(t,0)\hat B\,\rho(0)\right]
\nl&
\!+\!\int^t_{0}dt_2\int^{0}_{t_0}dt_1 \,{\rm Tr}\Big\{\hat A\,\Pi(t,t_2)
\Sigma_B(t_2-t_1)\rho(t_1)\Big\},
\end{align}
which is formally exact.  The relevant diagrams are shown schematically in Fig.\,\ref{figvertex}.
Here, $\Sigma_B $ is the self-energy of vertex corrections including
the operator $\hat  B$ and contain all the unseparable diagrams as well as 
the self-energy $\Sigma$ in the propagator  $\Pi $.
It is worth noting that \Eq{CABt-dia} has the same form as \Eq{CABth}(or \Eq{CABt}) where  
$\Sigma_B $ can be extracted and $\Sigma $
  is expressed explicitly in \Eq{selft} which has the same order of the magnitude to $\Sigma_B$.

We will now use this formalism to discuss the range of validity of the non-Markovian correlation function. As has been shown in
Ref.\,\onlinecite{Chr14104302}, in general non-Markovian effects are of the same order has higher order contractions in the 
self-energy. However, for the correlation function we have a well defined time scale and for short times scales, the combination of non-Markovian master equation and 
lowest order self-energy can be valid. 

Let us make the Taylor expansion of the time-derivative of the correlation function, i.e, 
$G^{\rm I} (t)\equiv\frac{d\la \hat A(t)\hat B(0)\ra^{\rm I} }{dt}$ in $H_{\s}$--interaction,
for small $t=0^+$,
\be
G^{\rm I} (t)=G^{\rm I} (0)+\frac{d G^{\rm I} (t)}{dt}\Big|_{t=0}\,t
+\frac{1}{2}\frac{d^2 G^{\rm I} (t)}{dt^2}\Big|_{t=0}\,t^2+\cdots.
\nonumber
\ee
Following the estimation of the order of magnitude of the
 self-energy in Ref.\,\onlinecite{Chr14104302},
we roughly get (see \App{app_Gt} for the detail),
\begin{align}\label{Gt}
&G^{\rm I} (t)\sim{\rm Tr} \big[\hat A f(Q,\hat B,\bar\rho)\big]\sum_{l}
\frac{g^{2l}}{\gamma^{2l-1}}
\nl&
+{\rm Tr}\Big\{\hat A \big[f(Q)\hat B\bar\rho+ f(Q,\hat B,\bar\rho)\big]\Big\}
\sum_{l}\frac{g^{2l}}{\gamma^{2l-2}}\,t
\nl&
+\frac{1}{2}{\rm Tr}\Big\{\hat A \big[f(Q)\hat B\bar\rho+ f(Q,\hat B,\bar\rho)\big]\Big\}
\sum_{l}\frac{g^{2l}}{\gamma^{2l-3}}\,t^2+\cdots,
\end{align}
where $\gamma$ is the smallest decay rate $\gamma_m$ of $C(t)$ in \Eq{cth}, $f(Q,\hat B,\bar\rho)$ 
and $f(Q)$ are just the formal expressions arising from $\Sigma_B$ and $\Sigma$, respectively.
Note that the magnitudes of $f(Q)\hat B\bar\rho$ and $f(Q,\hat B,\bar\rho)$ are nearly equal.
We conclude that the expansion is valid for time $t\lesssim1/\gamma$, even if only  the lowest order 
contributions $l=1$ are considered.

\section{Current fluctuations of the electron transport in quantum dots}
\label{thexa}

For the demonstration of characteristic non-Markovian effects, 
we consider electron transport through quantum dots (QDs)
contacted with two electrodes (left $L$ and right $R$).
This is a typical fermionic open quantum system where we will consider the non-Markovian effects 
in the spectrum in
high-frequency regime. \cite{Kac03035301,Eng04136602,Jin11053704}

The coupling Hamiltonian of \Eq{Hprime0} is specified by 
\be\label{HSQD}
H'=\sum_{ \mu}\left( a^\dg_{\mu }F_{ \mu}+F^\dg_{ \mu} a_{\mu }\right),
\ee
with the system operator $Q^{\pm}_\mu$ being $a^\dg_\mu$ ($a_\mu$) the creation (annihilation)
operator of an electron of the $\mu$th-level of the dot, and
the operator of the reservoirs
$F^\dg_{ \mu}=\sum_{\alpha=L,R} F^\dg_{\alpha \mu}$ with $ F^\dg_{\alpha \mu} =\sum_{k} t_{\alpha k\mu}
   c^{\dg}_{\alpha\mu k}$. The correlator of lead $\alpha$ in the frequency-domain of
   \Eq{FDT} is thus
  $C^{\pm}_{\alpha\mu}(\omega)
=\Gamma_{\alpha\mu}(\omega)f^{\pm}_\alpha(\omega)$
with the spectral density of the reservoir
$\Gamma_{\alpha\mu}(\omega)
=2\pi\sum_{k}t_{\alpha\mu k}t^\ast_{\alpha\mu k}\delta(\omega-\epsilon_k)$,
the Fermi function of $f^{+}_\alpha(\omega)=f_\alpha(\omega)=\frac{1}{1+\exp(\beta(\omega-\mu_\alpha))}$, $f^{-}_\alpha(\omega)=1-f_\alpha(\omega)$, and $\beta=1/{\rm k}_B T$ the inverse of the temperature. 
For the studied model, we consider $C^{(\pm)}_{\alpha\mu\nu} (t)=C^{(\pm)}_{\alpha\mu} (t)\delta_{\mu\nu}$
and symmetrical bias voltage $\mu_L=-\mu_R=eV/2$.

Assuming a Lorentzian spectrum centered around the
Fermi energy of the lead,\cite{Win9411040, Mac06085324,Kon95,Zhe08093016,Jin11053704,Jin14244111} 
$\Gamma_{\alpha\mu }(\omega)=
\frac{\Gamma_{\alpha\mu } {\rm w}^2}{(\omega-\mu_\alpha)^2+{\rm w}^2}$,
with high cut-off frequency 
 ${\rm w}=100\Gamma$ with $\Gamma=\sum_{\alpha\mu}\Gamma_{\alpha\mu}$,
it leads to \Eq{tcw0} [ \Eq{tcw}] \cite{Kon95,Jin14244111} $C^{\pm}_{\alpha\mu}(\omega)
=\Gamma_{\alpha\mu}(\omega)f^{\pm}_\alpha(\omega)\approx\Gamma_{\alpha\mu}f^{\pm}_\alpha(\omega)$
and
\begin{align}
\Lambda^{(\pm)}_{\alpha\mu }(\omega)
&=\frac{\Gamma_{\alpha\mu}}{\pi}
\Bigg\{{\rm Re}\left[\Psi\left(\frac{1}{2}
+i\frac{\beta(\omega-\mu_\alpha)}{2\pi}\right)\right]
&\nla
-\Psi\left(\frac{1}{2}+\frac{\beta {\rm w}_\alpha}{2\pi}\right)
\mp\pi\frac{\omega-\mu_\alpha}{{\rm w}_\alpha}\Bigg\},
\end{align}
with $\Psi(x)$ the Digamma function.

Here we focus on the investigation of the non-Markovian current fluctuations 
in the central dots, i.e., $\la \dot Q(t)\dot Q(0)\ra$ with 
$\dot Q=e \frac{d\hat N(t)}{dt}$ and $\hat N=\sum_\mu a^\dg_\mu a_\mu$. Since it satisfies the charge conservation of 
$\dot Q(t)=-\big[I_L(t)+I_R(t)\big]$, the corresponding spectrum is expected to be  closely related to
the noise spectrum of the transport current $I_\alpha(t)$ in the reservoir $\alpha$.  
Actually, the current-fluctuation spectrum in the central dot can be easily obtained 
in terms of the corresponding charge fluctuation defined by $S_N(\omega)={\cal F}\{\la \hat N(t)\hat N(0)\ra\}$,
via the relation of $S_{\rm c}(\omega)={\cal F}
\big[\la \dot Q(t)\dot Q(0)\ra\big]=e^2\omega^2 S_N(\omega)$.
The spectrum of the charge fluctuation $S_N(\omega)$
is given by the formula 
\Eq{CABw} with appropriately replacing $\hat A=\hat B=\hat N$.

\subsection{Single quantum dot}
\label{SQD}
Let us first study the simplest model 
of  electron transport through a spin-less one-level QD in the sequential tunneling regime,
 $\mu_L>\varepsilon>\mu_R$, described by the Hamiltonian $H_{\s}=\varepsilon a^\dg a$.
 Two states are considered in the dot, the empty state ($|0\ra$) and the single-electron occupied state ($|1\ra$),
 which leads to $a=|0\ra\la 1|$.
For the spectrum of the charge fluctuation $S_N(\omega)={\cal F}\{\la \hat N(t)\hat N(0)\ra\}$  calculated by \Eq{CABw},
the result agrees completely with that given in Ref.\,\onlinecite{Kac03035301} based on diagrammatic technique.

\begin{figure}
\centerline{\includegraphics*[width=1.0\columnwidth,angle=0]{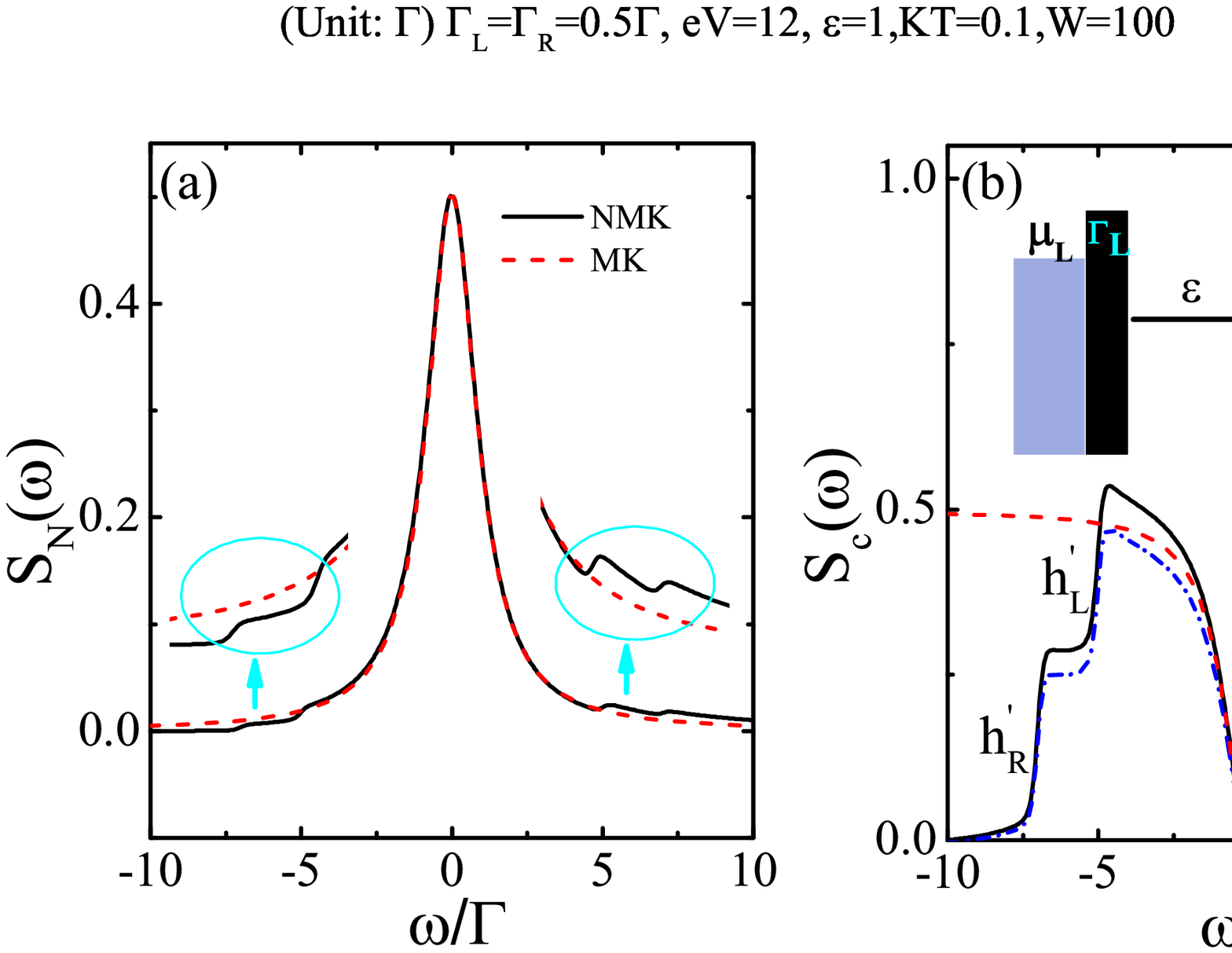}}  
\caption{(Color online)  
The spectra of charge and current fluctuations in (a) and (b), respectively,
for the single level quantum dot under the non-Markovian (black solid-line)
and Markovian (red dashed-line) treatments with symmetrical coupling $\Gamma_{ L}=\Gamma_R=0.5\Gamma$.
The blue dashed-dot-line in (b) is non-Markovian result for asymmetrical coupling $\Gamma_{ L}=2\Gamma_R$
for the confirmation of \Eq{heighp0} and \Eq{heighn0}.
The typical non- Markovian feature shows step at $\omega_{\alpha 0}=|\varepsilon-\mu_\alpha|$ in the spectra.
The other parameters are (in unit of $\Gamma=\Gamma_L+\Gamma_R$):  
$\varepsilon=1$, $k_BT=0.1$, and $\mu_L=-\mu_R=eV/2=6$.}.
  \label{SQD1}
\end{figure}

The numerical result is shown in Fig.\,\ref{SQD1}.
It depicts non-Markovian features (solid-line) compared to the Markovian results (dashed-line)
 (a) for charge fluctuation and (b) for current fluctuation, respectively.
 For low frequency regime at $\omega<\text{min}\{|\mu_\alpha-\varepsilon|\}$ corresponding to
 the long time limit,
 the results based on both non-Markovian and Markovian treatments are consistent
 due to the disappearance of non-Markovian effect. 
With increasing the frequency higher than the energy-resonance,
i.e., $\omega\gtrsim\omega_{\alpha 0}\equiv|\varepsilon-\mu_\alpha|$,
it enters the non-Markovian regime where
the non-Markovian characteristic occurs in the spectra, 
showing steps at $\omega\approx\omega_{\alpha 0}$ in
the current fluctuation spectrum (see Fig.\,\ref{SQD1} (b)).
This is consistent with the noise spectrum of the transport current 
through the reservoirs studied in Refs.\,\onlinecite{ Eng04136602,Jin11053704,Yan14115411}

The non-Markovian feature showing steps at $\omega_{\alpha 0}$ in $S_c(\omega)$ provides the information of the
energy structure in the central dot. 
The heights of the steps contain the information of the tunneling rate as 
demonstrated in the following.
For the studied single-level dot in the regime of $\mu_L>\varepsilon>\mu_R$, the stationary population of 
the empty and single-electron occupied states are,
$\bar\rho_{00}=\frac{\Gamma_R}{\Gamma}$ and $\bar\rho_{11}=\frac{\Gamma_L}{\Gamma}$, respectively.
Considering the spectrum in the positive frequency regime ($\omega>0$), 
it corresponds to the energy absorption processes.
Accordingly, when the dot is in the empty state $|0\ra$ with probability $\bar\rho_{00}$, 
the electrons in the right reservoir absorb the energy, i.e, $\mu_R+\omega=\varepsilon$ 
and tunnel to the dot, which 
leads to the step at $\omega_{R0}=|\varepsilon-\mu_R|$ with the height of $h_R\approx\bar\rho_{00} \Gamma_R=\frac{\Gamma^2_R}{\Gamma}$.
When the dot is in the occupied state $|1\ra$ with probability $\bar\rho_{11}$, 
the electrons in the dot absorb the energy, i.e, $\varepsilon+\omega=\mu_L$ and tunnel to the left reservoir, which 
leads to the step at $\omega_{L0}=|\mu_L-\varepsilon|$ with the height of $h_L\approx\bar\rho_{11} \Gamma_L=\frac{\Gamma^2_L}{\Gamma}$.
The ratio of the heights in the positive frequency regime 
thus is,
\be\label{heighp0}
h_L:h_R\approx{\Gamma^2_L}:{\Gamma^2_R};~~~(\omega>0).
\ee
Similarly, in the reversed regime ($\omega<0$) corresponding to the energy emission processes, 
we get $h'_L\approx\bar\rho_{00} \Gamma_L=\frac{\Gamma_L\Gamma_R}{\Gamma}$ and 
$h'_R\approx\bar\rho_{11} \Gamma_R=\frac{\Gamma_L\Gamma_R}{\Gamma}$,
which leads to 
\be\label{heighn0}
h'_L/h'_R\approx1;~~~(\omega<0),
\ee 
which is insensitive to the tunneling rate.

\begin{figure}
\centerline{\includegraphics*[width=0.8\columnwidth,angle=0]{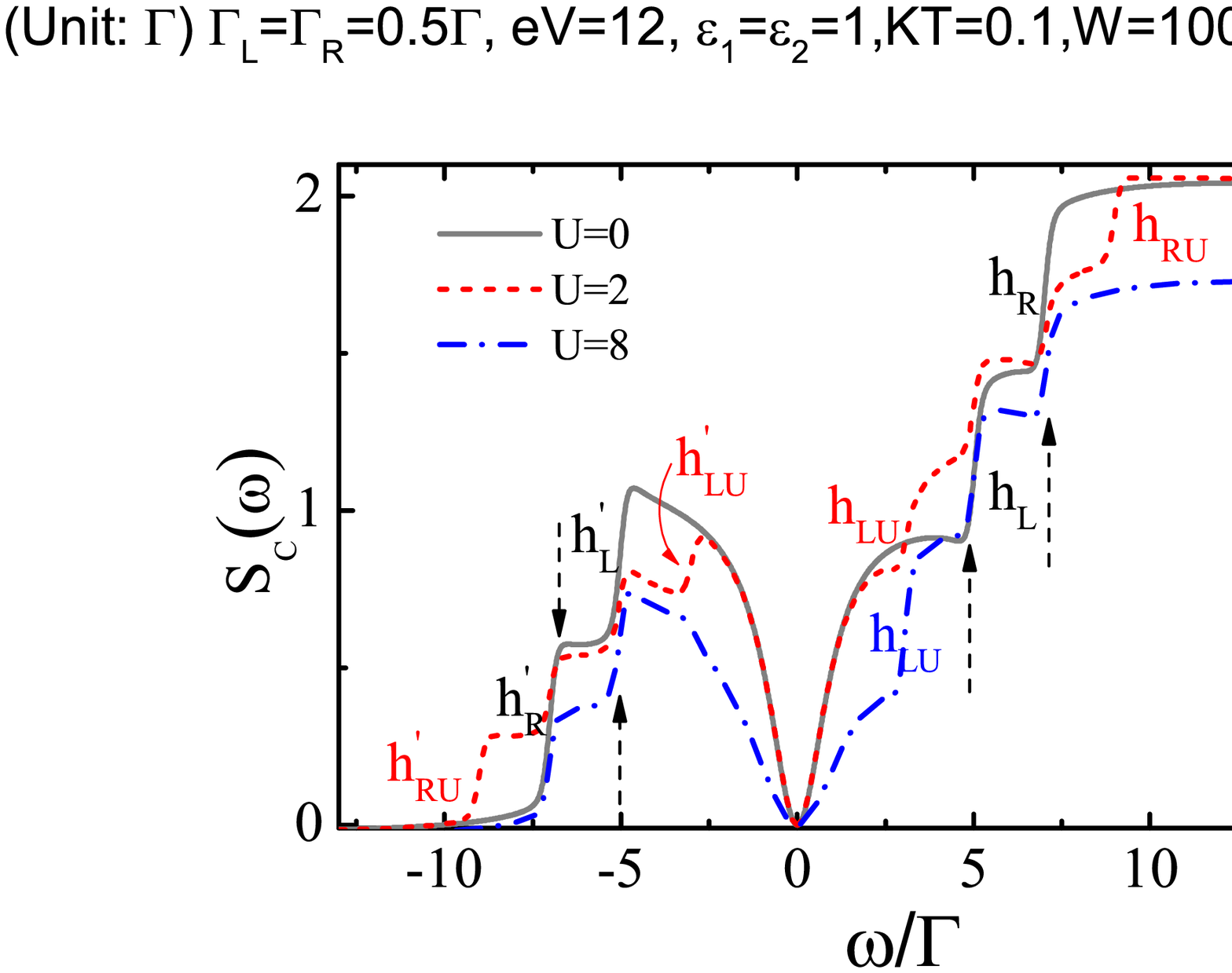}}  
\caption{(Color online)  
The current-fluctuation spectrum in the QD with different Coulomb interaction
for symmetrical coupling $\Gamma_{ L}=\Gamma_R=0.5\Gamma$.
It displays the non-Markovian steps occurring at not only $\omega_{\alpha 0}$ (denoted by the 
dashed arrows), but also $\omega_{\alpha U}=|\varepsilon+U-\mu_\alpha|$
induced by Coulomb interaction.
The other parameters are the same as in Fig.\,\ref{SQD1} }
  \label{SQD2}
\end{figure}

We further consider the single level in the dot
 with spin-dependence 
as described by the Hamiltonian,
$ H_{\s}= \sum_{\mu=\up,\down}\varepsilon_\mu a^\dg_\mu a_\mu 
   +U \hat n_\up \hat n_\down$ ,
where $\hat n_\mu=a^\dg_\mu a_\mu$ and $\hat N=\sum_{\mu=\up,\down}\hat n_\mu$. 
The involved states in the dot are $|0\ra$, $|\!\!\up\ra$, $|\!\!\down\ra$, and $|2\ra\equiv|\!\!\up\down\ra$
denoting the empty, two single--occupation spin states,
and the double--occupation spin--pair state, respectively.
In this state-basis, we have $a_{\down}=|0\ra\la \down|+|\up\ra\la 2|$
and $a_{\up}=|0\ra\la \up|-|\down\ra\la 2|$.
To demonstrate the Coulomb interaction effect more transparently,
we fix the dot level with spin-degeneracy,
$\varepsilon_\up=\varepsilon_\down=\varepsilon$, and consider spin-independent coupling strength,
$\Gamma_{\alpha \up}=\Gamma_{\alpha \down}=\Gamma_\alpha$, ($\alpha=L,R$).
The corresponding spectrum of the current fluctuations with different Coulomb interaction is shown in 
Fig.\,\ref{SQD2}. 
Besides the steps at the energy-resonance $\omega_{\alpha0}$, we also find the steps induced by Coulomb interaction 
  at $\omega_{\alpha U}\equiv|\varepsilon+U-\mu_\alpha|$. 
The different Coulomb interaction 
modifies the positions and the heights of the steps in the spectrum.
In the positive high frequency limit at $\omega>\text{max}\{\omega_{LU},\omega_{RU}\}$, 
the current fluctuation spectra nearly approach the same value due to 
the absorption of enough energy to open all the tunneling channels.

We identify the regimes as (i) weak $U$ with $\mu_L>\varepsilon,\varepsilon+U>\mu_R$ and 
(ii) strong $U$ with $\varepsilon+U>\mu_L>\varepsilon>\mu_R$.
In Fig.\,\ref{SQD2}, for $U=0$, 
the ratios of the heights of the steps both in the positive and negative parts
are the same as spinless single level. However, compared to Fig.\,\ref{SQD1},
the magnitude of the spectrum is doubled due to the two-energy levels (spin-up $|\up\ra$ and spin-down
$|\down\ra$) involved 
in the transport.
After the similar derivation in spinless single-level dot as demonstrated above,
 the ratios of the heights of the steps for the positive (denoted by $h$) and negative (denoted by $h'$) frequencies
  in the spectrum are given by, respectively,
\bsube
\begin{align}
h_{RU}: h_R: h_L: h_{LU}&=\Gamma_L\Gamma^2_R: \Gamma^3_R :\Gamma^2_L\Gamma_R :\Gamma^3_L,
\\
h'_{RU}: h'_R: h'_L: h'_{LU}&=\Gamma_L: \Gamma_R :\Gamma_R :\Gamma_L,
\end{align} 
\esube
for (i) weak $U$ (the short-dashed-line in Fig.\,\ref{SQD2}), and
\bsube
\begin{align}
h_{RU}: h_R: h_L: h_{LU}&=\Gamma_L\Gamma_R: \Gamma^2_R :\Gamma^2_L:\Gamma^2_L,
\\
 h'_R: h'_L&=1 :1,
\end{align} 
\esube
for (ii) strong $U$ (the dot-dashed-line in Fig.\,\ref{SQD2}). 
Since the stationary double occupation is not allowed for strong $U$,
there are no Coulomb-induced steps in the negative part of the spectrum.

\subsection{Coupled double quantum dots}

Now let us consider the electron transport through the system of
two coupled quantum dots described by the Hamiltonian
$
 H_{\s}= \varepsilon_{l}a^\dg_la_l + \varepsilon_{r}a^\dg_ra_r
   +U \hat n_l \hat n_r+\Omega\big(a^\dg_{l} a_{r}+a^\dg_{r} a_{l}\big)$,
where $U$ is the interdot Coulomb interaction, $\hat n_\mu=a^\dg_\mu a_\mu$ and $\hat N=\sum_{\mu=l,r}\hat n_\mu$. 
The involved states of the double dot are $|0\ra$ for the empty double dot, 
$|L\ra$ for the left dot occupied, $|R\ra$ 
for the right dot occupied, and $|2\ra\equiv|LR\ra$ for the two dots occupied.
Here, we assume at most one electron in each dot.
In this space, we have $a_{l}=|0\ra\la L|+|R\ra\la2|$
and $a_{r}=|0\ra\la R|-|L\ra\la 2|$.
The description of the involved states in this double dots 
is quite similar to that in the single dot with spin-dependence
studied above. However, the essential difference is that
the states of $|L\ra$ and $|R\ra$ are not the eigenstates of the
system Hamiltonian $H_{\s}$ which have the intrinsic coherent Rabi oscillation
demonstrated by the coherent coupling strength $\Omega$.
The corresponding Rabi frequency denoted by $\Delta$ is
the energy difference between the eigenstates ($\varepsilon_{\pm}$),
e.g., $\Delta=\varepsilon_{+}-\varepsilon_{-}=2\Omega$ for
the degenerate double-dots system considered here.

\begin{figure}
\centerline{\includegraphics*[width=1.0\columnwidth,angle=0]{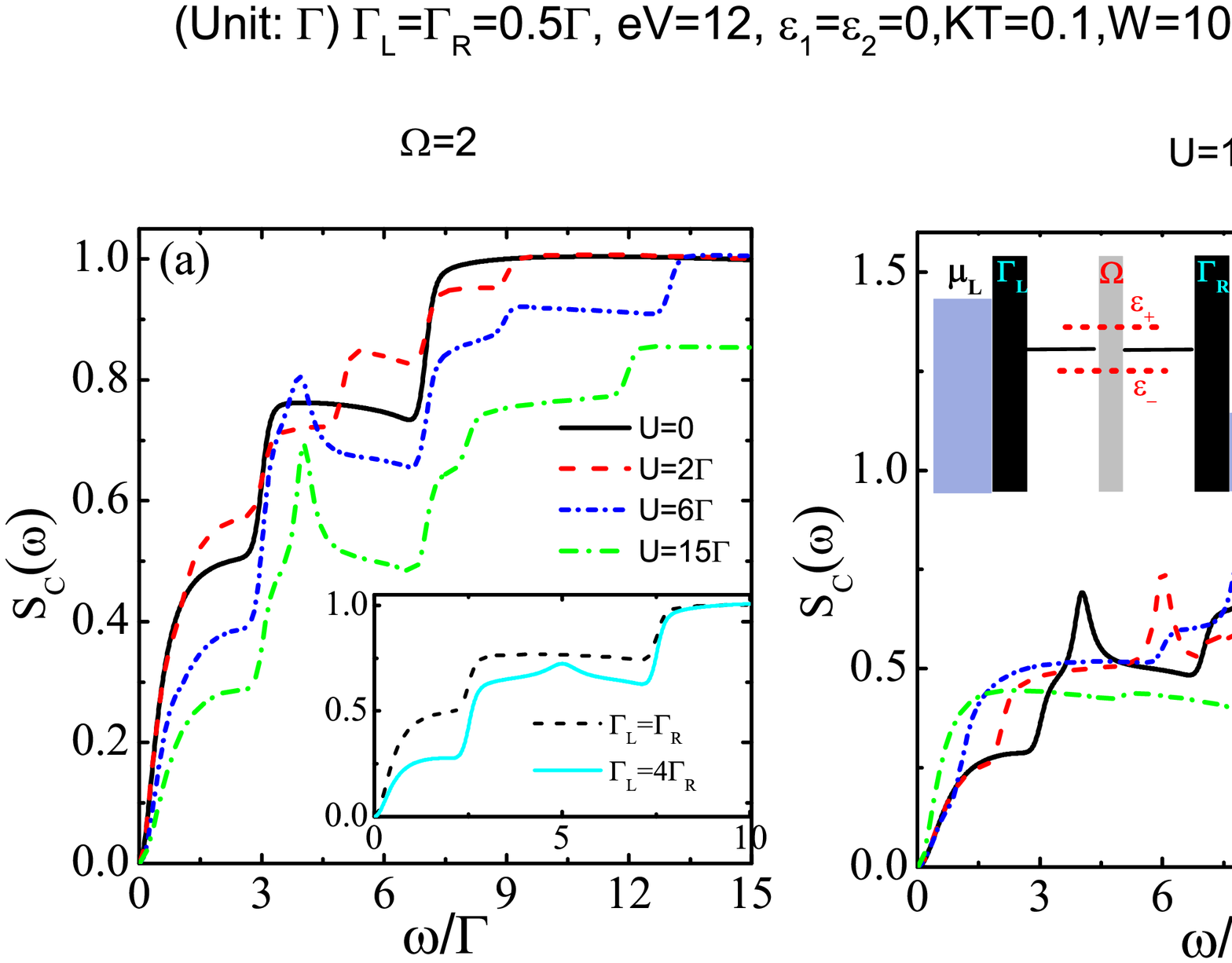}}
\caption{(Color online)  
 The charge-fluctuation spectrum in the coupled quantum dots (a) with different Coulomb interaction
for the interdot coupling $\Omega=2$ and (b) with different interdot coupling for strong Coulomb 
interaction $U=15$. The inset in (a) is the spectra comparison between the symmetrical and asymmetrical coupling for $\Omega=2.5$ and $U=0$.
The inset in (b) is the diagram of the transport setup for the electron transport through the coupled double quantum dots.
The other parameters are (in unit of $\Gamma=\Gamma_L+\Gamma_R$):  
$\varepsilon_l=\varepsilon_r=0$, $k_BT=0.1$, $\Gamma_L=\Gamma_R=0.5$, and $\mu_L=-\mu_R=eV/2=5$. }
  \label{CQD}
\end{figure}

The current-fluctuation spectrum for the coupled double dots is numerically displayed in Fig.\,\ref{CQD}.
Similar to the single quantum dot, the spectrum has the feature of a energy-resonance 
step at $\omega_{\alpha 0}=|\varepsilon_{\pm}-\mu_\alpha|$
and the Coulomb interaction induced step at $\omega_{\alpha U}=|\varepsilon_{\pm}+U-\mu_\alpha|$
as shown in Fig.\,\ref{CQD} (a). Here, we only illustrate the positive-frequency part of the spectrum 
due to the similar or less information involved in the negative-frequency regime. 
It is worth noting that the step behavior reflects the eigenstate 
energy structure of the dots, say, $\varepsilon_\pm$ rather than $\varepsilon_{l,r}$.
Besides the same step behavior as  in the single quantum dot, of particular interest is
 the signal of the coherent Rabi oscillation of the coupled double dots in the current-fluctuation spectrum.

Interestingly, the emergence of the coherent signal of the Rabi oscillation  
is nearly determined by
the strength of the Coulomb interaction $U$.
For weak Coulomb interaction in the regime of
 $\mu_L>\varepsilon_\pm,\varepsilon_\pm+U>\mu_R$, such as $U=0$ (solid-line) and $U=2\Gamma$ (dashed line)
 in Fig\,\ref{CQD} (a), no signal at the Rabi frequency $\omega=\Delta$ occurs in the spectrum.
While in the double-dot Coulomb blockade regime, either $\varepsilon_++U>\mu_L>\varepsilon_\pm,\varepsilon_-+U>\mu_R$ 
or $\varepsilon_\pm+U>\mu_L>\varepsilon_\pm>\mu_R$, the current-fluctuation spectrum always shows a peak
at the Rabi frequency $\omega=\Delta$ as shown in Fig\,\ref{CQD} (a) with $U=6\Gamma$ (short-dashed-dot-line) and 
$U=15\Gamma$ (dashed-dot-line).
Although the coherent signal peak appears for asymmetrical coupling with $\Gamma_L=4\Gamma_R$ 
shown in the inset of Fig\,\ref{CQD} (a), it is quite weak compared to that induced by Coulomb interaction.
This means the coherent Rabi oscillation information in the current-fluctuation spectrum is sensitive to
the dynamical blockade channel.
 This characteristic of Rabi coherence signal is also consistent with
  Markovian treatment studied in Ref.\,\onlinecite{Luo07085325}, where the symmetrical current fluctuation spectrum
  was considered.

Furthermore, by increasing the coherent coupling strength $\Omega$,
we find that the coherent signal of the Rabi oscillation is moved to the high-frequency regime 
with strong non-Markovian effect as shown in Fig\,\ref{CQD} (b).
Simultaneously, the peak of the coherent signal in the spectrum gradual 
 increases monotonically and sharply increases 
 at the resonance regime where the Rabi frequency approaches the bias voltage, i.e., $\Delta=2\Omega=eV$, as 
 shown in Fig.\,\ref{CQD} (c).
This arises from the interplay
between the Rabi resonance and the lead-dot tunneling
resonance, i.e., $\varepsilon_{\pm}(=\pm\Omega)=\mu_{L,R}(=\pm eV/2)$, combined with strong non-Markovian effect.
It may suggest that this resonant regime is good for
the observation of the coherent 
signal in the current-fluctuation spectrum experimentally. 
  Beyond the resonance regime, the system will enter into the cotunneling regime which is beyond the present approach
 and we have to recur to more advanced approaches such as the hierarchical equations of motion \cite{Jin08234703,Wan13035129,Jin15234108} and the real-time diagrammatic technique \cite{Kon95}
 for the consideration of higher-order contributions in the self-energy.

\section{Summary}
\label{sum}

In summary, using the frequency-dispersed technique by transforming the 
typical time-nonlocal master equation into equivalent
time-local equations-of-motion, we established an efficient 
formula for the two-time non-Markovian correlation function of arbitrary system operators
in open quantum systems.
The key to the calculation of the non-Markovian correlation function is 
how to restor an effective time-translation symmetry to the propagator.
We find that this corresponds to the vertex corrections
as further demonstrated by the real-time diagrammatic technique.
  The final result 
has an elegant structure and is as convinent to apply as the widely used quantum regression theorem
for the Markovian case.

  We applied the present method to study the 
  current-fluctuation spectra in the interacting single quantum dot and coupled double dots, 
  respectively, contacted by two electrodes.
  The typical non-Markovian effect have been demonstrated.
 We found that the non-Markovian step behavior in the current-fluctuation spectrum of the single quantum dot 
  is consistent with that in the noise spectrum of the transport current through the leads.
  The sharp peak of the coherent Rabi signal in the double dots occurs at the resonance regime where
  the eigenenergy levels are comparable to the chemical potential in the leads under the applied 
  bias voltage.
  From this current-fluctuation spectrum covering the full-frequency regime,
  the information of the energy structure of the quantum dots, the tunneling rate 
  as well as the Coulomb interaction and even the coherent Rabi signal can be extracted directly.

\acknowledgments
We acknowledge helpful discussions with G.~{Sch\"{o}n},
  X. Q. Li, and Y. J. Yan. J.S.J. acknowledges support from a fellowship
of the KIT, as well as support from the Program of HNUEYT,
the NNSF of China (Grant No. 11274085), NSF of Zhejiang Province (Grant No. LZ13A040002),
and Hangzhou Innovation Funds for Quantum Information and Quantum Optics.

\appendix
\section{The detail derivation of the two-time NMK-CF expressed in 
\Eq{CABt} and \Eq{CABw}}
\label{app_cf}
In this appendix, we give a detailed derivation of the non-Markovian
two-time correlation function given by the \Eq{CABt}.
For the calculation of the steady state as the initial condition of correlation function, the 
stationary solution of auxiliary density operators $\bar\phi^{\pm}(\omega)\equiv\phi^{\pm}(\omega,0)$ 
are given by \Eq{phi0}
\begin{align}\label{adotau}
\bar\phi^{\pm}(\omega)&=
-i\int^0_{-\infty}dt_1
e^{i({\cal L}_{\s}\mp\omega)t_1}
{\cal C}^{(\pm)}_Q(\omega)\rho(t_1).
 \end{align}
Updated with the initial condition of (c.f. \Eq{rhotau}),
$\hat{\vec\rho}(0)= \hat B\vec{\rho}(0)=\{ \hat B\bar\rho,\hat B \bar\phi^{+}(\omega),
\hat B \bar\phi^{-}(\omega)\}$ for \Eq{rho1t}, 
we get the formula of the auxiliary density operators
\begin{align}\label{app_ADOs}
\hat \phi^{\pm}(\omega,t)&=
e^{-i({\cal L}_{\s}\mp\omega)t}\hat B\bar\phi^{\pm}(\omega)
\nl&
-i\int^t_{0}dt_2
e^{-i({\cal L}_{\s}\mp\omega)(t-t_2)}{\cal C}^{(\pm)}_Q(\omega)\hat\rho(t_2).
   \end{align}
Inserting it into \Eq{rho0t} yields
\begin{align}\label{app_ME1}
 \dot{\hat\rho}(t) &= -i{\cal L}_{\s}\hat\rho(t)
  + \int_{0}^t\!\!{\mathrm d}\tau\,
   \Sigma(t-\tau)\hat\rho(\tau)
   \nl&
   -i\sum_{+,-}\int\frac{d\omega}{2\pi}
   [Q^{\mp},e^{-i({\cal L}_{\s}\mp\omega)t}\hat B\bar\phi^{\pm}(\omega)],
\end{align}
which has the solution
\begin{align}
 &{\hat\rho}(t) =\Pi(t,0)\hat B\bar{\rho}
 \nl&
  \! -\!i\sum_{+,-}\int^t_0\!\! dt_2\Pi(t,t_2)\!\int\!\!\frac{d\omega}{2\pi}
   [Q^{\mp},e^{-i({\cal L}_{\s}\mp\omega)t_2}\hat B\bar\phi^{\pm}(\omega)].
\nonumber
\end{align}
Here, we used the initial condition ${\hat\rho}(0)=\hat B{\rho}(0)=\hat B\bar{\rho}$.
With the use of \Eq{adotau} and the relation (c.f. \Eq{FDT} and \Eq{calCw})
\begin{align}\label{FDT2}
 \!\int\!\!\frac{d\omega}{2\pi}
   e^{\pm i \omega t} 
 {\cal C}^{(\pm)}_Q(\omega) 
={\cal C}^{(\pm)}_Q(t),
\end{align} 
based on \Eq{ABt0}, we finally get the NMK-CF expressed by \Eq{CABt}.
We could directly get the spectrum of the correlation function in the frequency domain by a Laplace transformation based on \Eq{CABt}.
An alternative way is that we first obtain the stationary solution of auxiliary density operators in \Eq{adotau} which reads ($\rho(t_1)=\Pi(t_1,t_0)\rho(t_0)=\bar\rho$
with $t_0\!\rightarrow\!-\infty$)
\begin{align}\label{ados}
\bar\phi^{\pm}(\omega) =\frac{{\cal C}^{(\pm)}_Q(\omega)\bar\rho}{\pm\omega-{\cal L}_{\s}+i0^+}.
\end{align}
Then, we transforme \Eq{app_ME1} into the frequency-domain with the relation
\begin{align}
\int \frac{d\omega'}{2\pi}\frac{\bar\phi^{\pm}(\omega) }{\omega-{\cal L}_{\s}\pm\omega'+i0^+}
= \!-\!
\frac{i}{\omega}(\widetilde{\cal C}^{(\pm)}_Q({\cal L},0)
\!-\!\widetilde{\cal C}^{(\pm)}_Q({\cal L},\omega))\bar\rho\big],
\nonumber
\end{align}
The result is
\begin{align}\label{hatrhow}
[\widetilde\Pi(\omega)]^{-1}\tilde{\hat \rho}(\omega) 
&=  \hat B\bar\rho\!-\!
\frac{i}{\omega} \!\sum_{+,-}\!\big[Q^{\mp}\!\!,\hat B
\nl&\quad \times
 (\widetilde{\cal C}^{(\pm)}_Q({\cal L},0)
\!-\!\widetilde{\cal C}^{(\pm)}_Q({\cal L},\omega))\bar\rho\big].
\end{align}
Finally, according to the first identity of \Eq{ABt0} which suggests
 $L [\la A(t)B(0)\ra]={\rm Tr}[\hat A \tilde{\hat \rho}(\omega)]$, we get the formula of the NMK-CF in the frequency-domain as expressed in \Eq{CABw}.

Similarly, the correlation function of $\la \hat A(0)\hat B(t)\ra$ is given by
 \begin{align}\label{CBAt}
&\la \hat A(0)\hat B(t)\ra \!=\!
{\rm Tr}\left[\hat B\,\Pi(t,0)(\bar\rho\hat A)\right]
\!-\!\sum_{+,-}\int^t_{0}\!\!dt_2\int^{0}_{-\infty}\!\!dt_1 
\times
\nl&
{\rm Tr}\Big\{\hat B\,\Pi(t,t_2)
\big[Q^{\mp},\Pi_0(t_2,0) \,\big(\Pi_0(0,t_1)
{\cal C}^{(\pm)}_Q(t_2-t_1)\bar\rho\hat A\big)\,\big]\Big\},
\end{align}
 in the time-domain and
\begin{align}\label{CBAw}
&L [\la A(t)B(0)\ra] 
={\rm Tr}[\hat B\, \widetilde\Pi(\omega)(\bar\rho\hat A)]\!-\!
\frac{i}{\omega}{\rm Tr}\Big\{\hat B\,\widetilde\Pi(\omega)
\nl&\quad\quad\quad\times
\!\sum_{+,-}\!\big[Q^{\mp}\!\!,  (\widetilde{\cal C}^{(\pm)}_Q({\cal L},0)
\!-\!\widetilde{\cal C}^{(\pm)}_Q({\cal L},\omega))\bar\rho\hat A\big] \Big\},
\end{align}
in the frequency-domain.

\section{The two-time correlation function for multiple coupling}
\label{app_Hmul}
For simplicity so far, we showed the derivation of the NMK-CF for a single-operator coupling formalism
as expressed in \Eq{Hprime0}. Realistically, the coupling Hamiltonian is given by the multiple-operator coupling formalism,
i.e., $H'=  Q^+_\mu F^-_\mu+{\rm H.c.}$ as we illustrated in \Sec{thexa}. 
The final formulas of \Eq{CABt} and \Eq{CABw} can be generalized by simply 
 replacing the operators $Q^{\pm}$ and ${\cal C}^{(\pm)}_Q$ with $Q^{\pm}_\mu$ and ${\cal C}^{(\pm)}_{Q_\mu}$,
 respectively, and further adding the sum of $\mu$.
Thus, the formalism presented in \Sec{thnmkME} and \Sec{thnmkcf} as well as \App{app_cf} are the same but adding
the symbol $\mu$ with the summation. Especially we pay attention on the 
final formula of NMK-CF in the time-domain of \Eq{CABt}, where
  ${\cal C}^{(\pm)}_Q$ expressed in \Eq{calCt}
 should be replaced by
 \be\label{calCtm}
 {\cal C}^{(\pm)}_{Q_\mu}(t-\tau) \bullet\!\equiv\!
 \sum_{\nu} \left[C^{(\pm)}_{\mu\nu}(t-\tau)Q^\pm_\nu\bullet-\bullet C^{(\mp)\ast}_{\mu\nu}(t-\tau)Q^\pm_\nu\right],
 \nonumber
 \ee
with 
   $ C^{(\pm)}_{\mu\nu} (t-\tau)
   = \la F^\pm_\mu (t) F^{\mp}_\nu (\tau)  \ra_{\B}$,
   and the counterpart in the frequency-domain expressed in \Eq{calCw} should be replaced by
\begin{align}
\widetilde {\cal C}^{(\pm)}_{Q_\mu}({\cal L},\omega)\bullet\!=\!
\sum_{\nu}\Big[\widetilde{C}^{(\pm)}_{\mu\nu}(\omega-{\cal L})Q^\pm_\nu\!\bullet
-\bullet\!\widetilde C^{(\mp)\ast}_{\mu\nu}({\cal L}-\omega)Q^\pm_\nu\Big],
\nonumber
\end{align}
with (c.f. \Eq{tcw0})
\begin{align}\label{tcw}
   \widetilde C^{(\pm)}_{\mu\nu} (\omega)
&=\int^\infty_{-\infty}\frac{d\omega'}{2\pi}
\frac{i}{\omega\pm\omega'+i0^+}C^{(\pm)}_{\mu\nu}(\omega')
\nl&=\frac{1}{2}\left[C^{(\pm)}_{\mu\nu}(\mp \omega)
 +i\Lambda^{(\pm)}_{\mu\nu}(\mp \omega)\right].
 \end{align}  
  For instance, the NMK-CF of \Eq{CABw} is generalized to
  \begin{align}\label{CABw2}
&L [\la A(t)B(0)\ra] 
={\rm Tr}[\hat A\, \widetilde\Pi(\omega)\hat B\bar\rho]\!-\!
\frac{i}{\omega}{\rm Tr}\Big\{\hat A\,\widetilde\Pi(\omega)
\nl&\quad\quad\times
\!\sum_{\mu,+,-}\!\big[Q^{\mp}_{\mu},\hat B (\widetilde{\cal C}^{(\pm)}_{Q_\mu}({\cal L},0)
\!-\!\widetilde{\cal C}^{(\pm)}_{Q_\mu}({\cal L},\omega))\bar\rho\big] \Big\}.
\end{align}

\section{Diagram description}
\label{app_dia}

\subsection{The derivation of formal exact two-time correlation function in \Eq{CABt-dia}}
\label{app_dia_CAB}
The two-time correlator of two operators $\hat A$ and $\hat B$ acting in the Hilbert space of the system of interest is given by
\be
  \langle \hat A(t)\hat B(0)\rangle
  \!=\!\text{Tr}\left[\rho_T(t_0)U(t_0,t)\hat A U(t,0) \hat B U(0,t_0)\right].
  \nonumber
\ee
The functions $U(t',t)$ are the unitary time evolutions 
$U(t',t)=\mathcal{T}\{e^{i\int_{t'}^t\text{d}t''H_{\rm tot}(t'')}\}$ and $\rho_T(t)$ 
is the density matrix of the total system. The operator $\mathcal{T}$ is 
the time ordering operator. We assume that we can write this density matrix for $t_0$ as 
a direct product of the reduced density matrices of the system and the bath, which 
is valid for the limit $t_0\!\rightarrow\! -\infty$,
\begin{align}
  \rho_T(t_0)=&\rho_B(t_0)\otimes\sum_{{s}{s}'}\rho_{{s}{s}'}(t_0)\ket{{s}}\bra{{s}'},
  \\
  \langle \hat A(t)\hat B(0)\rangle=&\text{Tr}\Big[\rho_B(t_0)\otimes
  \sum_{{s}{s}'}\rho_{{s}{s}'}(t_0)\ket{{s}}\bra{{s}'}\notag
  \\
  &U(t_0,t)\hat A U(t,0) \hat B U(0,t_0)\vphantom{\sum_{{s}{s}'}}\Big]
  \nl
  =&\sum_{{s},{s}'}\rho_{{s}{s}'}(t_0)\bra{{s}'}\Tr_B\left\{\rho_B(t_0)\right.\notag\\
  &\left.U(t_0,t)\hat A U(t,0) \hat B U(0,t_0)\right\}\ket{{s}},
\end{align}
where $\{|s\ra\}$ is a basis set for the central system.
By changing to the interaction picture (operators and states marked by tilde) the equation can be further simplified. Therefore, we introduce the Hamiltonian $H_0=H_S+H_B$ and the unitary time evolution $U_0(t',t)=\mathcal{T}\{e^{i\int_{t'}^t\text{d}t''H_0(t'')}\}$. The time evolution in the interaction picture is $\widetilde{U}(t',t)=U_0(t,t')U(t',t)=\mathcal{T}\{e^{i\int_{t'}^t\text{d}t''H'(t'')}\}$. 
An operator in the interaction picture is given by $\tilde{A}(t)=U^\dg_0(t,t_0)\hat AU_0(t,t_0)$ and an expansion
 of the exponential functions $\tilde{U}(t,t')$ yields an expansion in the couplings $H'$
\begin{align}
  \tilde{U}&(t_0,t)\tilde{A}(t)\tilde{U}(t,0)\tilde{B}(0)\tilde{U}(0,t_0)=\tilde{A}(t)\tilde{B}(0)\notag
  \\
  &+\int_0^t\text{d}t_1\int_0^{t_1}\text{d}t_2 \tilde{H}'(t_1)\tilde{A}(t)\tilde{H}'(t_2)\tilde{B}(0)\notag
  \\
  &-\int_0^t\text{d}t_1\int_0^{t_1}\text{d}t_2 \tilde{H}'(t_2)\tilde{H}'(t_1)\tilde{A}(t)\tilde{B}(0)+\cdots,
  \\  
  \langle A&(t)B(0)\rangle^{\rm I}\!=\!\!\begin{tikzpicture}[anchor=base,baseline=8pt]
    \coordinate (A) at (0,0);
    \coordinate (B) at (1,0);
    \coordinate (C) at (0,0.7);
    \coordinate (D) at (1,0.7);
    \coordinate[label=right:{\small $\tilde{A}(t)$}] (E) at (1.2,0.35);
    \coordinate[label=below:{\small $\tilde{B}(0)$}] (F) at (0.25,0);
    \draw[line width=1.0pt] (A) -- (B);
    \draw[line width=1.0pt] (B) to [bend right=45](E);
    \draw[line width=1.0pt] (E) to [bend right=45](D);
    \draw[line width=1.0pt] (D) -- (C);
    \fill (F) circle (2pt);
    \fill (E) circle (2pt);
    \end{tikzpicture}\!+\!
    \begin{tikzpicture}[anchor=base,baseline=8pt]
    \coordinate (A) at (0,0);
    \coordinate (B) at (1,0);
    \coordinate (C) at (0,0.7);
    \coordinate (D) at (1,0.7);
    \coordinate[label=right:{\small $\tilde{A}(t)$}] (E) at (1.2,0.35);
    \coordinate[label=below:{\small $\tilde{B}(0)$}] (F) at (0.25,0);
    \coordinate (G) at (0.5,0);
    \coordinate (H) at (0.8,0.7);
    \draw[line width=1.0pt] (A) -- (B);
    \draw[line width=1.0pt] (B) to [bend right=45](E);
    \draw[line width=1.0pt] (E) to [bend right=45](D);
    \draw[line width=1.0pt] (D) -- (C);
    \fill (F) circle (2pt);
    \fill (E) circle (2pt);
    \fill (G) circle (2pt);
    \fill (H) circle (2pt);
    \end{tikzpicture}\notag\\
    &\quad\quad\quad\!+\!
    \begin{tikzpicture}[anchor=base,baseline=8pt]
    \coordinate (A) at (0,0);
    \coordinate (B) at (1,0);
    \coordinate (C) at (0,0.7);
    \coordinate (D) at (1,0.7);
    \coordinate[label=right:{\small $\tilde{A}(t)$}] (E) at (1.2,0.35);
    \coordinate[label=below:{\small $\tilde{B}(0)$}] (F) at (0.25,0);
    \coordinate (G) at (0.8,0);
    \coordinate (H) at (0.5,0.7);
    \draw[line width=1.0pt] (A) -- (B);
    \draw[line width=1.0pt] (B) to [bend right=45](E);
    \draw[line width=1.0pt] (E) to [bend right=45](D);
    \draw[line width=1.0pt] (D) -- (C);
    \fill (F) circle (2pt);
    \fill (E) circle (2pt);
    \fill (G) circle (2pt);
    \fill (H) circle (2pt);
    \end{tikzpicture}+
    \begin{tikzpicture}[anchor=base,baseline=8pt]
    \coordinate (A) at (0,0);
    \coordinate (B) at (1,0);
    \coordinate (C) at (0,0.7);
    \coordinate (D) at (1,0.7);
    \coordinate[label=right:{\small $\tilde{A}(t)$}] (E) at (1.2,0.35);
    \coordinate[label=below:{\small $\tilde{B}(0)$}] (F) at (0.25,0);
    \coordinate (G) at (0.8,0.7);
    \coordinate (H) at (0.5,0.7);
    \draw[line width=1.0pt] (A) -- (B);
    \draw[line width=1.0pt] (B) to [bend right=45](E);
    \draw[line width=1.0pt] (E) to [bend right=45](D);
    \draw[line width=1.0pt] (D) -- (C);
    \fill (F) circle (2pt);
    \fill (E) circle (2pt);
    \fill (G) circle (2pt);
    \fill (H) circle (2pt);
    \end{tikzpicture}\!+\cdots,
     \end{align} 
 which leads to
\begin{align}
   \langle \hat A(t)\hat B(0)\rangle^{\rm I}
   =\sum_{{s},{s}'}\rho_{{s}{s}'}(t_0)\prod_{{s}{s}'}(t_0,\tilde{A}(t),\tilde{B}(0)).
\end{align}
Each dot in a diagram denotes a coupling Hamiltonian $H'$. 
The $\prod_{{s}{s}'}(t_0,\tilde{A}(t),\tilde{B}(0))$ superoperator is 
the full time evolution of the density matrix including the two operators $\hat A$ and $\hat B$. 
By using Wicks theorem the trace over the bath decays into two point functions represented 
by a contraction between the dots. So we can write 
\begin{align}
    \begin{tikzpicture}[anchor=base,baseline=8pt]
    \coordinate (A) at (0,0);
    \coordinate (B) at (1,0);
    \coordinate (C) at (0,0.7);
    \coordinate (D) at (1,0.7);
    \coordinate[label=right:{\small $\hat{A}(t)$}] (E) at (1.2,0.35);
    \coordinate[label=below:{\small $\hat{B}(0)$}] (F) at (0.25,0);
    \coordinate (G) at (0.1,0.7);
    \coordinate (H) at (0.1,0);
    \coordinate[label=above:{$\prod$}] (K) at (0.5,0.1);
    \draw[line width=1.0pt] (A) -- (B);
    \draw[line width=1.0pt] (B) to [bend right=45](E);
    \draw[line width=1.0pt] (E) to [bend right=45](D);
    \draw[line width=1.0pt] (D) -- (C);
    \draw[line width=1.0pt] (G) -- (H);
    \draw[line width=1.0pt] (B) -- (D);
    \fill (F) circle (2pt);
    \fill (E) circle (2pt);
    \end{tikzpicture}\hspace{-0.2cm}=\hspace{-0.2cm}
    \begin{tikzpicture}[anchor=base,baseline=8pt]
    \coordinate (A) at (0,0);
    \coordinate (B) at (2,0);
    \coordinate (C) at (0,0.7);
    \coordinate (D) at (2,0.7);
    \coordinate[label=right:{\small $\tilde{A}(t)$}] (E) at (2.2,0.35);
    \coordinate[label=below:{\small $\tilde{B}(0)$}] (F) at (1,0);
    \coordinate (G) at (0.2,0.7);
    \coordinate (H) at (0.2,0);
    \coordinate (I) at (1,0.7);
    \coordinate[label=above:{$ \widetilde\prod$}] (K) at (0.5,0.05);
    \coordinate[label=above:{$ \widetilde\prod$}] (J) at (1.5,0.05);
    \draw[line width=1.0pt] (A) -- (B);
    \draw[line width=1.0pt] (B) to [bend right=45](E);
    \draw[line width=1.0pt] (E) to [bend right=45](D);
    \draw[line width=1.0pt] (D) -- (C);
    \draw[line width=1.0pt] (G) -- (H);
    \draw[line width=1.0pt] (F) -- (I);
    \draw[line width=1.0pt] (B) -- (D);
    \fill (F) circle (2pt);
    \fill (E) circle (2pt);
    \end{tikzpicture}\hspace{-0.2cm}+\hspace{-0.2cm}
    \begin{tikzpicture}[anchor=base,baseline=8pt]
    \coordinate (A) at (0,0);
    \coordinate (B) at (2,0);
    \coordinate (C) at (0,0.7);
    \coordinate (D) at (2,0.7);
    \coordinate[label=right:{\small $\tilde{A}(t)$}] (E) at (2.2,0.35);
    \coordinate[label=below:{\small $\tilde{B}(0)$}] (F) at (1,0);
    \coordinate (G) at (0.2,0.7);
    \coordinate (H) at (0.2,0);
    \coordinate (L) at (0.7,0.7);
    \coordinate (M) at (0.7,0);
    \coordinate (N) at (1.4,0.7);
    \coordinate (O) at (1.4,0);
    \coordinate[label=above:{$ \widetilde\prod$}] (K) at (0.45,0.05);
    \coordinate[label=above:{$ \widetilde\prod$}] (J) at (1.7,0.05);
    \coordinate[label=above:{$ \widetilde\sum_{\B}$}] (P) at (1,0.05);
    \draw[line width=1.0pt] (A) -- (B);
    \draw[line width=1.0pt] (B) to [bend right=45](E);
    \draw[line width=1.0pt] (E) to [bend right=45](D);
    \draw[line width=1.0pt] (D) -- (C);
    \draw[line width=1.0pt] (G) -- (H);
    \draw[line width=1.0pt] (L) -- (M);
    \draw[line width=1.0pt] (N) -- (O);
    \draw[line width=1.0pt] (B) -- (D);
    \fill (F) circle (2pt);
    \fill (E) circle (2pt);
    \end{tikzpicture}
    \label{eCut}
\end{align}
Here, $\sum_{\B}$ which is the self-energy for vertex corrections contains all the unseparable diagrams 
and \Eq{eCut} 
is still exact. We can rewrite this equation in algebraic form as 
\begin{align}
 \langle \hat A&(t)\hat B(0)\rangle^{\rm I}=\Tr\Big\{\tilde{A}(t)\widetilde\Pi(t,0)\widetilde{B}(0)\tilde\Pi(0,t_0)\rho(t_0)\notag\\
 &+\int_0^t\text{d}t_2 \int_{t_0}^0\text{d}t_1\tilde{A}(t)\widetilde\Pi(t,t_2)\widetilde\Sigma_{B}(t_2,t_1)\widetilde\Pi(t_1,t_0)\rho(t_0)
 \Big\}.
 \label{e2tcor}
\end{align}
Using the relation of $\rho(0)=\Pi(0,t_0)\rho(t_0)$ and $\rho(t_1)=\Pi(t_1,t_0)\rho(t_0)$,
\Eq{e2tcor} immediately recast to 
\Eq{CABt-dia} in the Schrodinger picture.

\subsection{The derivation for \Eq{Gt}}
\label{app_Gt}

We introduce $\varrho_{\B}(t)\equiv\Pi(t,0)\hat\rho_{\B}(0)$ with $\varrho_{\B}(0)=\hat B\,\bar\rho$,
and the auxiliary density operator describing the
vertex contribution,
\begin{align}\label{rhoGB}
\varrho_{\G \B}(t)=\int^t_{0}dt_2\int^{0}_{t_0}dt_1  \Pi(t,t_2)
\Sigma_{\B}(t_2-t_1)\bar\rho,
\end{align}
with $\varrho_{\G \B}(0)=0$ apparently.
Since we are interested in the discussion of the order of the magnitude roughly,
it is more convenient in the $H_{\s}$-interaction picture.
The corresponding time-derivation equations are
\bsube\label{varrho1t}
\begin{align}
\dot{\tilde\varrho}_{ \B}(t)&=  \int_{0}^t\!\!{\mathrm d}\tau
   \widetilde\Sigma(t-\tau)\tilde\varrho_{ \B}(\tau),
   \\
\dot{\tilde\varrho}_{\G \B}(t)&=  \int_{0}^t\!\!{\mathrm d}\tau
    \widetilde\Sigma(t-\tau)\tilde\varrho_{\G \B}(\tau)
   +\tilde f_{\B}(t),
\end{align}
\esube
with $\tilde f_{\B}(t) =\int^{\infty}_{t}dt_1  \widetilde\Sigma_{\B}(t_1)\bar\rho$.
Then the time-derivation of the two-time correlation function (c.f. \Eq{e2tcor}) reads
\begin{align}\label{Gt0}
G^{\rm I}(t)\!\equiv\!\frac{d\la \hat A(t)\hat B(0)\ra^{\rm I}}{dt} 
\!=\!
{\rm Tr}\left\{\hat A\,\big[\dot{\ti\varrho}_{\B}(t)+\dot{\ti\varrho}_{\G \B}(t)\big]\right\}.
\end{align}
The Taylor expansion of the time-derivative of the correlation function
for small $t=0^+$ is
\be
G^{\rm I} (t)=G^{\rm I} (0)+\frac{d G^{\rm I} (t)}{dt}\Big|_{t=0}\,t
+\frac{1}{2}\frac{d^2 G^{\rm I} (t)}{dt^2}\Big|_{t=0}\,t^2+\cdots.
\nl
\ee
With \Eq{varrho1t}, we get
\begin{align}\label{Gt2}
G^{\rm I} (t)&= {\rm Tr} \big[\hat A\ti f_B(0)\big]
+{\rm Tr}\Big\{\hat A
\big[ \widetilde\Sigma(0)\hat B+ \widetilde\Sigma_B(0)\big]\bar\rho\Big\}\,t
\nl&\quad
+\frac{1}{2}{\rm Tr}\Big\{\hat A
\big[\dot{\widetilde\Sigma}(0)\hat B+\dot{\widetilde\Sigma}_B(0)\big]\bar\rho\Big\}\,t^2+\cdots.
\end{align}
Using the estimation of the order of magnitude presented in  Ref.\,\onlinecite{Chr14104302}, we roughly get  
\begin{align}\label{para_order}
\ti f_B(0)  
&\sim\sum_{l}
\frac{g^{2l}}{\gamma^{2l-1}}f(Q,\hat B,\bar\rho),
\nl
 \widetilde\Sigma_B(0)\bar\rho&\sim\sum_{l}
\frac{g^{2l}}{\gamma^{2l-2}}f(Q,\hat B,\bar\rho),
\nl
\dot{\widetilde\Sigma}_B(0)\bar\rho&\sim\sum_{l}
\frac{g^{2l}}{\gamma^{2l-3}}f(Q,\hat B,\bar\rho),
\nl
 \widetilde\Sigma(0)&\sim\sum_{l}
\frac{g^{2l}}{\gamma^{2l-2}}f(Q),
\nl
\dot{\widetilde\Sigma}(0)&\sim\sum_{l}
\frac{g^{2l}}{\gamma^{2l-3}}f(Q),
\end{align}
where $\gamma$ is the minimum decay rate $\gamma$ of $C(t)$ in \Eq{cth}, $f(Q,\hat B,\bar\rho)$ 
and $f(Q)$ are just the formal expression arising from $\Sigma_B$ and $\Sigma$, respectively.
Inserting \Eq{para_order} into \Eq{Gt2}, we finally get \Eq{Gt}.

\end{document}